\documentclass[12pt]{article}
\usepackage[margin=.8in]{geometry}
\usepackage[utf8]{inputenc}
\usepackage{amsmath}
\usepackage{booktabs}
\usepackage{dirtytalk}
\usepackage{hyperref}
\usepackage{microtype}
\usepackage{natbib}
\usepackage{txfonts}
\usepackage{tipa}

\newcommand{\concept}[1]{\textit{#1}}
\newcommand{\abbr}[1]{#1}

\begin{document}

\title{Group-matching algorithms for subjects and items}
\author{Géza Kiss, Kyle Gorman, and Jan P.H. van Santen}
\date{}

\maketitle

\abstract{
We consider the problem of constructing matched groups such that the
resulting groups are statistically similar with respect to their average
values for multiple covariates. This group-matching problem arises in
many cases, including quasi-experimental and observational studies in
which subjects or items are sampled from pre-existing groups, scenarios
in which traditional pair-matching approaches may be inappropriate. We
consider the case in which one is provided with an existing sample and
iteratively eliminates samples so that the groups \say{match} according
to arbitrary statistically-defined criteria. This problem is NP-hard.
However, using artificial and real-world data sets, we show that
heuristics implemented by the \texttt{ldamatch} package produce
high-quality matches.
}

\section{Introduction}
\label{introduction}

When studying the effect of group differences on target variables, it is
essential to control the effects of other confounding variables
associated with the dependent and independent variables under
consideration. One can deal with potential confounding covariates at
various points during a study \citep[p.~55]{Szatmari2004}. First, one
can design the study to minimize confounds using a randomized controlled
trial or stratified sampling. Such designs also minimize the effect of
unobserved independent variables. Secondly, one can perform analysis on
a subsample chosen so that the groups are statistically similar. We
refer to this subsampling technique as \concept{matching}. Third and
finally, covariates can be dealt with in the analysis phase using
statistical techniques such as ANCOVA or multivariate regression. In
this section, we motivate the use of matching. In subsequent sections,
we present methods for selecting matched groups, implemented in the R
package \texttt{ldamatch}.

Subject matching is a common strategy in quasi-experimental or
observational studies in which subjects' assignment to group is
pre-existing or otherwise not under experimental control. For instance,
an experimenter might use post-hoc subject matching to compare diseased
or impaired humans to those without a disease or impairment. While less
common, matching of items or stimuli grouped into \say{conditions} or
treatments can also be used. For example, an experimenter studying
lexical processing might wish to balance different word lists for
frequency, length, and so on.

Some researchers have questioned whether experimenters ought to use
matching at all, particularly in the study of developmental disorders
\citep[e.g.,][]{Jarrold2004}. Covariate analysis is recommended as an
alternative to matching when, due to a large difference between group
characteristics, matching would distort the groups so much that they
would no longer be faithful representations of the populations from
which they were sampled \citep{Seltzer2004,Tager-Flusberg2004}. For
example, \citet{Jarrold2004}, argue that matching subjects with autism
from a wide range of intellectual abilities on IQ could easily reduce
the autism group to only those with very high IQ. They also observe
(p.~84) that
\say{matching groups on more than one criterion is often extremely difficult and,
even if possible,
will involve such a degree of selectivity that the generalizability of the findings will be reduced considerably.}
Thus \citeauthor{Jarrold2004} conclude that matching on multiple
variables is non-trivial, and may produce non-representative subsamples.
On the other hand, if there is little residual variance associated with
the covariates targeted, one would suppose that subject matching could
be applied, and would likely require few subjects to be removed.

It should be noted that one can combine matching with covariate analysis
during statistical inference. The matching framework does not make any
specific assumptions about the relationship between variables of
interest, and statistical techniques like ANCOVA or multivariate
regression can control for any group differences remaining after
matching is applied \citep{Tager-Flusberg2004}.
\citet[p.~349]{Blackford2009} argue that matching has other advantages,
including that
\say{matching produces effect-size estimates with smaller variance than covariate adjustment,
analyses on matched data are more robust,
and matching can control for more confounders than covariate adjustment,
for a given sample size.}

Perhaps the best-known matching technique is \concept{pair-matching}
as proposed by \citep{Rubin1973,Rosenbaum1983}. The idea behind this approach is that
if we could know the outcome in a person's life both if they receive or
do not receive a certain treatment, then we could easily measure the
effect of that treatment. This can be approximated as follows. First we
identify pairs of subjects that are similar on all relevant variables
before treatment. One of the subjects then receives the treatment, and
we measure changes in all outcome variables. One can assume that any
difference arising between them later is the effect of the treatment. If
this procedure is repeated for many pairs of subjects, one can calculate
a quantity known as the
\concept{Average Treatment Effect for the Treated Subjects}
(\abbr{ATT}), an estimate of the effect of a treatment on the dependent
variable or variables. The goal of pair-matching is thus to pair control
and treated subjects such that the pairs are very similar with respect
to the variables used for matching before receiving the treatment.
One-to-one, one-to-N, and M-to-N pairing can all be used in this
setting, since one can just as well compare outcome averages when
multiple subjects are selected. As a side effect, the selection of
well-matched subject pairs results in a good balance between groups of
the selected subjects overall \citep{Gu1993}. The notion of
\say{treatment} is quite general: it might include actual drug
treatment, partaking in an intervention or change of habit, or even the
prior presence or absence of a disease or disorder. Thus the above
approach can be used in experimental, quasi-experimental, and
observational studies. Several implementations for pair-matching are
available, including the R packages \texttt{Optmatch}
\citep{Hansen2007}, \texttt{MatchIt} \citep{Ho2011}, and
\texttt{Matching} \citep{Berkeley2011}, all of which support two-group
pair-matching; the \texttt{twang} R package \citep{Ridgeway2014}
supports pair-matching for two or more groups.

Pair-matching is generally performed based on
\concept{propensity scores}, the estimated probability of a subject
being assigned to the treatment group. The use of propensity scores
results in an unbiased estimate of the treatment effect so long as
outcomes are independent of assignment to treatment conditional on
pre-treatment covariates \citep[p.~151]{Dehejia2002}.
\citet{Blackford2009} gives several other arguments for propensity score
matching. For instance, more subjects can be preserved than when
matching on multiple covariates, so the resulting groups are more
similar to the overall population and less bias is introduced.
Furthermore, it is much easier to match on one variable than on multiple
dimensions \citep{Dehejia2002,Smith2005}. However, when matching on
multiple covariates two subjects may have very similar propensity scores
even when their individual characteristics differ widely. In other
words, propensity score matching does not guarantee that the paired
subjects will be impressionistically similar. For this reason, some
researchers use propensity scores but also require paired subjects to
have quite similar covariate values according to certain predefined
conditions. Propensity scores must be derived from pre-treatment
variables \citep[e.g.,][]{Rubin1991}, particularly those likely to be
both related to group assignment and relevant for the dependent
variables; however, they may not be variables caused by group assignment
or later outcomes \citep[p.~98]{Blackford2006}. Propensity scores can be
estimated in a number of ways, including a logistic regression model fit
to pre-treatment covariates.

Unfortunately, we do not always have enough information for estimating
propensity scores before treatment, and using variables that are
themselves affected by the treatment or variables that are not related
to the outcome results in \concept{overmatching}, matching that is
superfluous or erroneous and which harms the statistical efficiency or
the validity of the study; \citealp[e.g.,][]{Rothman2008}). For example,
if we view the presence of Autism Spectrum Disorder (ASD) in a subject
as a \say{treatment effect}, it is not clear what one could use as a
\say{pre-treatment} variable. There are many factors known to be related
to autism risk, including gender, the number of affected siblings, birth
order, genetic issues such as de novo mutations, and the presence of
certain environmental toxins and air pollution during gestation
\citep[e.g.,][]{Chaste2012}, but many of these risk factors are
difficult or impossible to measure long after birth. When we do not have
that information, it may be tempting to use whatever is available to
estimate propensity scores. Note however that autism is often comorbid
with intellectual disability (in about 50--70\% of all cases; see
\citealp{Matson2009}), so we cannot exclude the possibility that
intellectual (dis)ability is affected by having autism (although it
could be the other way around, or both could be the result of a common
underlying neurological condition). The situation is similar for
Specific Language Impairment (SLI), another developmental disorder: IQ
is generally lower in children with SLI, but it is not clear whether or
not the condition causes the lower average IQ. In either case, using IQ
to estimate propensity scores may produce overmatching.

Post-hoc matching may still be appropriate in those cases where
propensity scores cannot be estimated due to a lack of suitable
pre-treatment covariates. While it does not guarantee an unbiased
estimate of the treatment effect, post-hoc matching does eliminate the
possibility that an observed difference is due simply to group
differences in the matched covariates. For example, for children with a
neurodevelopmental disorder, the \say{post-treatment} covariates might
be cognitive and/or language abilities measured using standard
instruments. Performing analyses on data matched with respect to these
covariates does not ensure that the actual effect of treatment (here,
presence or absence of a development disorder) is properly measured, but
it does ensure that any group differences in the outcome variables are
unmediated by group differences in these covariates.

In other cases, including those for which propensity scores cannot be
estimated, one may alternatively use another type of matching we refer
to as \concept{group-matching}. Group-matching ensures that the
statistical distribution of covariates is similar between groups of
subjects---instead of at the individual level, as in pair-matching---by
ensuring that the groups are not significantly different with respect to
the covariates at some given \(\alpha\)-level; \citet{Mervis2004}, for
example, propose using \(\alpha > 0.5\) or \(\alpha > 0.2\). The
statistics used for matching are often based on comparison of group
means---\citet{Rubin1973} refers to this as
\concept{mean-matching}---but one may also wish to take other
statistical properties, like variance, into account. \citet{Rubin1973}
examines both matching approaches and reports group-matching works well
when the dependent variable is linearly related to the matching
variables. Moreover, matched groups can easily be analyzed using
standard statistical techniques such as mixed-effect linear models
\citep{PinheiroBates2000,Bates2005,Bates2010}. \citet[p.~37]{Shaked2004}
reports that
\say{larger effect sizes were yielded when participants\ldots were matched on a group basis,
rather than on a one-to-one basis, with the comparison participants.}

The computational implementation of group-matching has received very
little attention. For instance, in a 2004 special issue of the Journal
of Autism and Developmental Disorders, focused on the question of
subject matching, implementation was discussed only by one paper:
\citet{Mervis2004} describe a procedure for matching on one variable,
and it is not clear whether the procedure was implemented
computationally. We are unaware of prior work on group-matching with
multiple variables, a situation in which propensity scores may also be
difficult to estimate or to use.

In this paper, we introduce and evaluate multiple algorithms for
group-matching that uses complex matching criteria involving one or more
covariates. We have made available our implementation of those
algorithms to the research community via the \texttt{ldamatch} R
package. Rather than illustrating the use of this thoroughly-documented
package via code snippets, we describe these algorithms and evaluate
their effectiveness and performance by applying them to artificial and
real-world data sets.

\section{Problem statement}
\label{problem-statement}

Let us suppose that we have \(G\) groups comprising a total of \(N\)
subjects, with group membership for the subjects indicated by the vector
\(g_1,~\ldots,~g_N\). Each subject also has an associated covariate
vector \(\overline{c}_{i}, i=1~\ldots~N\). We wish to find an optimal
subset of subjects denoted by the boolean indicator variables
\(s_1,~\ldots,~s_N\) subject to some criteria. Without loss of
generality, let us assume that the criteria are a set of one or more
statistical tests that typify the similarity of the covariate
distribution between the groups, and possibly also constrain the
expected group size proportions or bound the maximum number of subjects
that can be removed, overall or from particular groups. The statistical
tests \(t_j, j=1~\ldots~T\) defining these criteria are assumed to be
functions from \(g_i\), \(\overline{c}_i\), and \(s_i\),
\(i=1~\ldots~N\) to \(p\)-values \(\in [0, 1]\). We say that the
difference between groups is non-significant if the \(p\)-values from
each statistical test is above a pre-specified threshold \(\alpha_j\);
for example, \(\alpha_j\) might be \(0.2\) or \(0.5\). Note that
matching does not consider the dependent variables targeted by the
larger study.

Our goal is to find solutions that meet the following expectations, in
decreasing order of importance. First, we want to preserve as many
subjects as possible. Second, we may have certain preferences related to
group membership: either want to maintain the ratio of the group sizes
close to a given ratio (such as the original group size ratio), or we
may prefer to keep subjects in certain groups more than in others; this
can be implemented by specifying a precedence among the groups. Third,
we want to minimize differences between the groups. For each criterion
we track the ratio of the actual \(p\)-value to the desired threshold
\(p\)-value. To combine this ratio across all criteria, we take the
mininum:

\begin{equation*}
r = \min_{j = 1~\ldots~T} \frac{p_j}{\alpha_j} .
\end{equation*}

\noindent
We maximize \(r\), halting when \(r \ge 1\). When comparing possible
matched subject configurations, these metrics can be used, in the above
priority ordering, to select a unique solution. Solutions for which the
above metrics are identical are assumed to be equivalent.

The above problem formulation is an optimization problem in which one
seeks optimal values for boolean variables \(s_1,~\ldots,~s_N\) such
that \(s_i\) is true if and only if subject \(s_i\) is preserved by the
matching. Were we to evaluate all possible combinations, there would
thus be \(2^N\) such solutions, making the evaluation exponential in the
number of subjects \(N\). We are interested in the solutions with the
highest total number of subjects, so we can introduce a lower bound
\(n\) defined such that we do not consider solutions which entail the removal of more than \(n\) subjects. There are

\begin{equation*}
\sum_{i=0}^{n}{N \choose i} = 1~+~N~+~{N \choose 2}~+~\ldots~+~{N \choose n}
\end{equation*}

\noindent
solutions that satisfy this constraint. Clearly, exhaustive search
quickly becomes intractable if many subjects must be removed to find a
solution. Furthermore, exclusion of any remaining part of the search
space may result in a suboptimal solution.

The matching problem is a discrete optimization problem, more
specifically an integer program---though not necessarily an integer
linear program---which are known to be NP-hard
\citep[p.~8f.]{Nemhauser1999}. Integer linear programs can be solved in
polynomial time using a relaxation to linear program and then mapping
the real values found onto integers. However, a solution found via this
relaxation may not be optimal, or even feasible. Furthermore, this
relaxation limits the types of statistical criteria that may be used.
Here we take a different approach. First, we propose heuristics which
search a well-defined subset of the hypothesis space. The following two
sections describe and evaluate these heuristics.

\section{Matching algorithms}
\label{matching-algorithms}

Below we describe several search strategies that evaluate a subset of
all possible subject configurations denoted by \(s_i, i=1~\ldots~N\),
with the goal of finding one which satisfies all statistical criteria at
the required \(\alpha\)-levels. We walk the search space with the aim of
optimizing the measures described above. Here we briefly describe these
algorithms, referring to them by the names used in \texttt{ldamatch} R
package.

\subsection{Random search (\texttt{random})}
\label{random-search-random}

This algorithm randomly samples the search space for a given number of
iterations choosing the subjects to keep randomly according to the
binomial distribution, gradually decreasing the expected value of their
count from \(N\) (the number of subjects) to \(G\) (the number of
groups). The search stops after the specified number of iterations \(I\)
and yields the best solutions found. This is a non-deterministic
algorithm with \(O(I \cdot T)\) running time, that is, it depends only
on the required number of iterations and on how long it takes to
evaluate the statistical tests for any particular subject configuration.

\subsection{Test-statistic search (\texttt{heuristic2})}
\label{test-statistic-based-search-heuristic2}

This is a constructive algorithm: it constructs a solution in a series
of steps, always choosing the next step such that it brings us nearer to
a solution (see for example \citealp{Genova2011}). The basic idea is
that we use the value of \(r\) (which indicates how well the groups are
matched) to decide which way to proceed when walking the search space,
that is, which subject to remove next to attain the largest improvement
in the target criteria. In every step, it calculates \(r\) after
removing each remaining subject in turn, and then removes the subject
that results in the highest \(r\). If multiple configurations with the
same number of subjects meet our criteria, we take advantage of the
other metrics to rank them and choose the first one. This is a
deterministic algorithm, with a computational complexity of
\(O(N^2 \cdot T)\); that is, it is quadratic in the number of subjects.
This approach was used in \citet{vanSanten2010Computational} for
selecting a matched subset of subjects for analysis.

\subsection{Test-statistic search with lookahead (\texttt{heuristic3} and \texttt{heuristic4})}
\label{test-statistic-based-search-with-lookahead-heuristic3-and-heuristic4}

Intuitively an issue with the \texttt{heuristic2} algorithm is that it
is not able to proceed toward the global optimum when the step in that
direction results in a local drop in \(r\). For example, when two
subjects need to be removed with extreme covariate values on the
opposite ends of the scale, the removal of either subject makes the
group balance worse.

We addressed the above issue in the \texttt{heuristic3} and
\texttt{heuristic4} algorithms by looking ahead several steps during the
search. Lookahead has been utilized for various problems, including
vehicle routing problems \citep{Atkinson1994}, decision tree induction
\citep{DongKothari2001}, and composition of finite-state automata
\citep{Allauzen2010}, but we are unaware of it being applied to the
matching problem. These two heuristics differ in how they decide on
their next step.

Both algorithms first identify one or more sets of \(L\) subjects,
denoted here by \(S\), whose removal results in the biggest improvement
(more than one set if they are equivalent on our metrics), then remove
one subject from those sets. Note that this way it is possible to reach
one of the best sets found \(L\) steps later, but it does not commit to
removing \(L\) particular subjects at this point, as it may find a still
better combination in later steps. In \texttt{heuristic3}, removal
decisions are made solely on the basis of \(r\), whereas
\texttt{heuristic4} prefers to eliminate a subject which is a candidate
for removal in the highest number of subject sets. Thus
\texttt{heuristic4} makes as minimal a future commitment as possible.

More formally, the algorithm for choosing the next subject for removal
in \texttt{heuristic3} is as follows.

\begin{enumerate}
\item Initialize \(l\) as \(L\) (the lookahead) and \(C\) as \(S\) (the candidates sets for removal, having the highest \(r\) values, consisting of \(L\) subjects each).
\item If \(l = 1\), choose the subject to be removed randomly from \(C\) and exit.
\item Decrement \(l\) (i.e., \(l \leftarrow l - 1\)).
\item Let \(S_l\) be all subject subsets of size \(l\) from \(C\).
\item For each subject subset in \(S_l\), calculate \(r\) with its subjects removed.
\item Let \(C\) be the subject sets from \(S_l\) with the highest \(r\) value.
\item Go to step 2.
\end{enumerate}

The algorithm for choosing the next subject for removal in
\texttt{heuristic4} is as follows:

\begin{enumerate}
\item Count the number of times each subject occurs in \(S\) and keep the ones with the largest count as candidates.
\item If more than one candidate remains, calculate \(r\) for each one and keep the ones with the highest \(r\).
\item If still more than one candidate remains, choose one of them randomly.
\end{enumerate}

Both algorithms are non-deterministic (as they choose randomly among
seemingly equivalent options), and follow a depth-first search strategy.
Their complexity is \(O(N^{L+1} \cdot T)\) where \(N\) is the number of
subjects and \(L\) is the degree of lookahead. Both algorithms are
equivalent to \texttt{heuristic2} when \(L = 1\).

The \texttt{ldamatch} implementation makes it possible to remove several
subjects without recalculating the \(r\) values. It can be
time-consuming to calculate the \(r\) value for such data sets, and
these values may not change much after removing one subject,
particularly when working with very large groups. The user can limit the
maximum number of subjects removed in each step (specifying it as a
number or as a percentage of the remaining subjects) and when the
algorithm should revert to calculating \(r\) after each step (e.g., when
\(r\) reaches some specified value). This lazy recomputation technique
does not change assymptotic complexity, but in practice, it speeds up
search substantially without major degradation in quality. As seen
below, allowing the removal of \(100\) subjects before recomputing \(r\)
may reduce the running time by more than an order of magnitude.

\subsection{Exhaustive search (\texttt{exhaustive})}
\label{exhaustive-search-exhaustive}

While exhaustive search is simple, it is often infeasible for large
problems. We implemented an exhaustive search algorithm that is feasible
when only a few subjects need to be removed to reach well-matched
subject groups. We can estimate an upper bound on the number of subjects
that need to be removed using the heuristics, and based on that number,
we can estimate the maximum running time of exhaustive search. Having
this algorithm at our disposal can give us the optimal solution in
certain cases and can enable us to assess how well other approaches
fare. Consider a fictious example when the exhaustive search is feasible
for a seemingly complex matching problem. Given two groups, each
containing 20 subjects, and assuming that the computer can process 1,000
out of the \(2^{20+20}\) subject configurations per second (which is
over \(1.099 \cdot 10^{12}\) configurations), evaluating all cases would
take over 34 years. However, if a heuristic finds a solution that meets
the matching criteria by removing five subjects, then we estimate that
exhaustive search will complete in 13 minutes or less, which makes
running an exhaustive search feasible. If it turns out during the
process that the optimal solution requires the removal of only three
subjects, then it will finish in less than 11 seconds.

The \texttt{ldamatch} package implements exhaustive search using a
breadth-first strategy. Our criteria rank candidate solutions first
based on the total number of subjects retained, and second by favoring
smaller divergence from the desired group proportions. Deviation from
the desired group proportions are measured using the Kullback-Leibler
(K--L) divergence \citep[see e.g.][]{CoverThomas2006}. When multiple
solutions are available with the same size and K--L divergence, we favor
those with higher values of \(r\).

\section{Evaluation}
\label{evaluation}

An optimal matching can be found using exhaustive search, but this is
not feasible for large or complex problems. Moreover, even when
exhaustive search is feasible, it may be preferable to find an
approximate solution if one can do so in a fraction of the time. The
heuristics generally find a solution, but we do not know if they do
indeed work in all cases and when they do, how much worse these
solutions are compared to the optimal one (i.e., their approximation
bounds).

\subsection{Procedures}
\label{procedures}

Our goal is to characterize these heuristics: their execution time, and
the quality of the solutions, comparing to an optimal solution found via
exhaustive search when possible. With this knowledge, we should be able
to choose a suitable algorithm for the task at hand and to estimate the
quality of approximation with respect to the optimal solution. For this
purpose, we evaluate the algorithms on several data sets: artificial
\say{subjects} from multivariate normal distributions with diverse
parameters, and real data sets, containing either a small group of
subjects or a large number of items. We analyze the results to gain an understanding of the pros and cons of the proposed algorithms.

\subsubsection{The matching process}
\label{the-matching-process}

Since every matching algorithm has its own respective drawbacks and
benefits, we do not commit to using any one of them, but rather make use
of several and select the best output. We use exhaustive search when
heuristics indicate that it will be feasible.

We hypothesized that heuristics would find better solutions when
stricter criteria are used. For example, they sometime preserves more
subjects when matching not just means, but overall distributional shape
\citep{Facon2011}, or when required to keep all subjects from one of the
groups. This may be because stricter criteria can help to guide them
into a direction that will prove to be globally optimal. In other words,
additional criteria may help the system to make better local decisions
and thus to attain better global decisions. We can take advantage of
this by matching the groups using both more and less strict criteria and
comparing the outputs.

\subsubsection{Computational resources}
\label{computational-resources}

We ran all evaluations in parallel using the Slurm Workload Manager to
launch jobs across a large, heterogeneous computing cluster. The
\texttt{exhaustive}, \texttt{heuristic3}, and \texttt{heuristic4}
algorithms were run in multithreaded mode. The nodes of the cluster were
commodity x86--64 machines running Ubuntu Linux, with 24 CPUs and 24
logical cores per node. CPU frequency varied from 1600 to 3000 MHz.

\subsection{Evaluation on synthetic data}
\label{evaluation-on-synthetic-data}

We first synthesize sets of items to be matched by randomly sampling
multivariate distributions. We sample from multivariate normal
distributions used to approximate an enormous number of
naturally-occurring phenomena. This allows us to evaluate the algorithms
for various parametric distributions without the difficulties of
collecting real data sets with desired properties.

\subsubsection{Data generation}
\label{data-generation}

Each synthetic data set is generated as follows. A random covariance
matrix is generated using the \texttt{genPositiveDefMat} function of the
\texttt{clusterGeneration} R package, then two random samples were drawn
from that distribution: a larger sample and a smaller sample with a
shifted mean. We refer to the former sample as \say{basic items} and the
latter as \say{intruders}. We combine the two samples into one data set
with the expectation that the matching procedure will identify and
exclude many intruders.

We created \(36\) parameter sets based on the following meta-parameters:

\begin{itemize}
\item number of all items: 100, 150, or 200 (to have a variety of configurations to evaluate).
\item total number of covariates: 2 to 4.
\item number of covariates that differ between the basic items and the intruders: 2 to 4 (e.g., 2 of 3)
\item number of intruders: 10
\item group sizes: 50\% of items in either group
\item means of each covariate: a random number drawn from a uniform distribution over \((1.0, 2.0)\).
\item difference between the means of the basic items and the intruders: between \(0.5\) and \(1.0\) times the diagonal of the covariance matrix
\item variances of each covariate: the mean multiplied by a random number from a uniform distribution over \((1.0, 10.0)\)
\item \(p\)-value for the basic subjects: between \(0.2\) and \(0.5\)
\item \(p\)-value for all subjects: \(< 0.1\)
\end{itemize}

We set the \(p\)-value threshold for the matching algorithms to \(0.2\).
This choice of \(p\)-values is intended to bias the algorithms toward
excluding intruders. However, when there is distributional overlap
between the basic items and the intruders, it is possible to achieve a
match without excluding all intruders; it is also possible that basic
items will be excluded.

We created \(5\) random data sets for each of the \(36\) collections of
parameters, and ran each method for each data set: the proposed
heuristics, and four results from 1, 10, 100, or 1,000 runs of random
search. The criteria used are the \(p\)-values from Welch's \(t\)-test
and the Anderson--Darling test, enforcing a match both on means and
distributional shapes.

\subsubsection{Metrics}
\label{metrics}

Our evaluation metrics are the following:

\begin{itemize}
\item the percentage of excluded items and of excluded intruders
\item the balanced divergence, the Kullback--Leibler divergence of the group ratios from the expected ratios (i.e., those for the original group ratios)
\item criteria \(p\)-values after matching
\item (\say{wall clock}) execution time
\end{itemize}

\subsubsection{Results}
\label{results-simulation}

We find (see \autoref{matching.art_data}) that the random algorithm
identified matched groups about 30\% of the time when it was run a
single time, excluding over 97\% of the items on average. When
preserving the best of 10 runs, the success rate increases to 99\%, and
100\% for 100 or 1000 runs. At the same time, the percentage of excluded
items decreases dramatically. This is significantly worse than the
results from the heuristics, though running times are comparable.

The heuristics succeed in finding a solution in every case. The
percentage of excluded subjects did not differ between them for the task
at hand: they generally find similar solutions, but running times
increases exponentially as lookahead is increased. A lookahead of 3 does
not improve match quality, but does find significantly more solutions
that satisfied our criteria. The heuristics do not quite meet
performance of exhaustive search. Exhaustive search is not only feasible
for most of these data sets (over 80\% of the time when we allowed it to
run for up to five days), but even takes lesss time to run than
heuristics with a lookahead of 3.

In summary, the heuristics are preferable to the random algorithms, at
least for this simulated data. However, they still do not always reach
the performance of the exhaustive algorithm, and can take very long to
run when a large lookahead is used.

\begin{table}
\centering
\begin{tabular}[c]{@{}lrrrrrr@{}}
\toprule
algorithm & \# solutions & \% E.~items & \% E.~intruders & BD & \(p\) & time \\
\midrule
\(r_1\) & 0 & 97 & 100 & .00 & .52 & \(< 1\) minute \\
\(r_{10}\) & 1 & 37 & 40 & .84 & .28 & \(< 1\) minute \\
\(r_{100}\) & 1 & 9 & 10 & .24 & .24 & \(< 1\) minute \\
\(r_{1000}\) & 1 & 4 & 10 & .06 & .22 & \(< 1\) minute \\
\(h_2\) & 1 & 2 & 10 & .12 & .22 & 1 minute \\
\(h_3\) (\(\lambda = 1\)) & 17 & 2 & 10 & .05 & .21 & 2 minutes \\
\(h_4\) (\(\lambda = 1\)) & 18 & 2 & 11 & .05 & .21 & 2 minutes \\
\(h_3\) (\(\lambda = 2\)) & 78 & 2 & 10 & .02 & .21 & 2 hours \\
\(h_4\) (\(\lambda = 2\)) & 74 & 2 & 10 & .02 & .21 & 2 hours \\
\(h_3\) (\(\lambda = 3\)) & 216 & 2 & 7 & .02 & .21 & 45 hours \\
\(h_4\) (\(\lambda = 3\)) & 221 & 2 & 7 & .02 & .21 & 45 hours \\
exhaustive & 2 & 2 & 10 & .02 & .22 & 3 hours \\
\bottomrule
\end{tabular}
\caption{Results from running the matching algorithms on synthetic data
sets; the median result across all data sets is reported. \# solutions:
number of unique solutions; \% E.~items: the percentage of excluded
items; \% E.~intruducers: the percentage of excluded items which were
intruders; BD: balanced divergence; \(p\): the \(p\)-value from Welch's
\(t\)-test or the Anderson-Darling test (whichever is smaller) for the
matched groups; \(r_{n}\): the best of \(n\) random algorithm runs;
\(h_2\): \texttt{heuristic2}; \(h_3\): \texttt{heuristic3}; \(h_4\):
\texttt{heuristic4}; \(\lambda\): lookahead.}
\label{matching.art_data}
\end{table}

\subsection{Matching subjects on neurocognitive measures}
\label{matching-subjects-on-neurocognitive-measures}

In this section, we describe the application of the matching algorithms
to the subject pool of the CSLU ERPA Corpus
\citep{Gorman2015Automated,Gorman2016,MacFarlane2017}. This enables us
to evaluate the algorithms on real-world data. First we summarize our
goals for the matching task, then we describe a process for finding a
solution. Finally we present some properties of the results and compare
the algorithms. For more information on this data set and the matching
problem it poses, see \citealt{Kiss2017}.

\subsubsection{Matching criteria}
\label{subject-matching-criteria}

Our goal is to find four sets of subjects from four groups such that
various group pairs are matched on a separate set of covariates. The
four groups are:

\begin{itemize}
\item ALN: Autism with normal language
\item ALI: Autism with language impairment
\item SLI: Specific Language Impairment
\item TD: typical development
\end{itemize}

\noindent
All groups are to be matched on age. Furthermore, particular pairs of
groups are to be matched on the following covariates:

\begin{itemize}
\item SLI and ALI: performance IQ (PIQ) and verbal IQ (VIQ)
\item ALI and ALN: ADOS score \citep{Lord2000Autism}
\item ALN and TD: PIQ and VIQ
\end{itemize}

There are 113 subjects in all, comprised of 43 TD, 25 ALN, 26 ALI, and
19 SLI diagnostic cases. We wish to keep all subjects with SLI (the
smallest group in our corpus), and as many as possible from the ALI,
ALN, and TD groups, in decreasing order of preference. The criteria
require a two-tailed \(\alpha > .2\) on Welch's \(t\)-test and,
optionally, the Anderson--Darling test.

A pilot experiment revealed that matching each pair of groups
independently gave different subsets of subjects for each pair. This was
considered undesirable as we want to find just one set of subjects that
meets all of the criteria above while also optimizing the group size and
other metrics. We address this problem by matching the groups
simultaneously, globally optimizing the solution. The basic idea of this
approach is to create one complex set of matching criteria that contains
everything we require in the final solution, and then use that to work
with all subjects from all groups at the same time. We therefore
implemented infrastructure components that allow us to:

\begin{enumerate}
\item specify complex sets of matching criteria, including various criterion functions and thresholds for various sets of groups
\item calculate \(r\) (the minimum \(p\)-value--threshold ratio) based on a combination of test statistics
\item calculate evaluation metrics for how well a set of subjects suits the matching criteria
\item enforce an ordering among the possible subject configurations; that is, being able to decide which one of two subject sets has better evaluation metrics
\end{enumerate}

\subsubsection{Results}
\label{results-ERPA}

Using \texttt{ldamatch}, we apply the simultaneous optimization approach
as outlined above to the CSLU ERPA ADOS Corpus. Results are given in
\autoref{matching.ERPA.conc_viq_stn.ados_ca.t_test.0.2.all}, which uses
the Welch's \(t\)-test criterion, and
\autoref{matching.ERPA.conc_viq_stn.conc_piq_stn.ados_sev.ados_ca.t_ad_test.0.2.all},
which combines the \(t\)-test and the Anderson--Darling test. Exhaustive
search is infeasible for this problem. In contrast, the \texttt{random}
algorithm practically always finds a solution, but the quality of the
solution is much worse than that obtained with heuristics. Regarding
these heuristics, we can see that there is not a large difference
between the algorithms in the number of subjects retained. Of the 113
candidate subjects, the best solutions preserved at most one more than
the others. However, the heuristic solutions are differentiated by the
degree to which they are able to preserve the desired group size
proportions and how many equivalent solutions they produce.

\begin{table}
\centering
\begin{tabular}[c]{@{}lrrr@{}}
\toprule
algorithm & \# excluded & \# solutions & time \\
\midrule
\(r_{1000}\) & 54 & 1 & \(< 1\) minute \\
\(h_2\) & 18 & 1 & \(< 1\) minute \\
\(h_3\) (\(\lambda = 1\)) & 17 & 16 & \(< 1\) minute \\
\(h_4\) (\(\lambda = 1\)) & 17 & 12 & \(< 1\) minute \\
\(h_3\) (\(\lambda = 2\)) & 17 & 96 & 22 minutes \\
\(h_4\) (\(\lambda = 2\)) & 17 & 165 & 22 minutes \\
\(h_3\) (\(\lambda = 3\)) & 17 & 39 & 9 hours \\
\(h_4\) (\(\lambda = 3\)) & 17 & 468 & 10 hours \\
\bottomrule
\end{tabular}
\caption{Results from ten replications of matching subjects from the
ERPA data set on verbal IQ and chronological age using Welch's
\(t\)-test at \(\alpha = .2\). \# excluded: the number of subjects
excluded (best of ten); \# solutions: number of unique solutions
(greatest of ten); time: wall clock execution time; \(r_{1000}\): the
best of \(1,000\) runs of the \texttt{random} algorithm; \(h_2\):
\texttt{heuristic2}; \(h_3\): \texttt{heuristic3}; \(h_4\):
\texttt{heuristic4}; \(\lambda\): lookahead; \# solutions: number of
unique solutions.}
\label{matching.ERPA.conc_viq_stn.ados_ca.t_test.0.2.all}
\end{table}

\begin{table}
\centering
\begin{tabular}[c]{@{}lrrr@{}}
\toprule
algorithm & \# excluded & \# solutions & time \\
\midrule
\(r_{1000}\) & 59 & 1 & \(< 1\) minute \\
\(h_2\) & 27 & 1 & 10 minutes \\
\(h_3\) (\(\lambda = 1\)) & 21 & 11 & 14 minutes \\
\(h_4\) (\(\lambda = 1\)) & 21 & 11 & 14 minutes \\
\(h_3\) (\(\lambda = 2\)) & 19 & 35 & 8 hours \\
\(h_4\) (\(\lambda = 2\)) & 19 & 30 & 8 hours \\
\(h_3\) (\(\lambda = 3\)) & 19 & 55 & 212 hours \\
\(h_4\) (\(\lambda = 3\)) & 19 & 40 & 211 hours \\
\bottomrule
\end{tabular}
\caption{Results from ten replications of matching subjects from the
ERPA data set on verbal IQ, performance IQ, ADOS severity score, and
chronological age using Welch's \(t\)-test and the Anderson--Darling
test at \(\alpha = .2\). \# excluded: the number of subjects excluded
(best of ten); \# solutions: number of unique solutions (greatest of
ten); time: wall clock execution time; \(r_{1000}\): the best of
\(1,000\) runs of the \texttt{random} algorithm; \(h_2\):
\texttt{heuristic2}; \(h_3\): \texttt{heuristic3}; \(h_4\):
\texttt{heuristic4}; \(\lambda\): lookahead; \# solutions: number of
unique solutions.}
\label{matching.ERPA.conc_viq_stn.conc_piq_stn.ados_sev.ados_ca.t_ad_test.0.2.all}
\end{table}

\subsection{Matching verbs on frequency} 
\label{matching-verbs-on-frequency}

Much of the prior work on matching is focused on quasi-experimental,
between-subjects studies in which differences between two or more groups
of human subjects is one of the primary sources of variance. However,
the matching procedures described above can just as easily be applied to
linguistic \say{items}---such as words---used as stimuli in experimental
or observational designs. As an example, we illustrate the matching of
regular and irregular English verbs on frequency and related attributes.
The regular verbs include all those verbs which form their past tense by
the addition of the suffix spelled \emph{-ed} and pronounced as [t],
[d] or [\textschwa d], depending on the final segment of the
verb stem. In contrast, the irregular verbs either undergo stem changes
in the past tense (e.g., \emph{become}-\emph{became},
\emph{leave}-\emph{left}, \emph{think}-\emph{thought}) or have a past
tense forms that are identical to their present tense forms (e.g.,
\emph{cut}, \emph{hit}, \emph{shut}). The regular verbs greatly
outnumber the irregular verbs, and it is well-known that the irregular
verbs have higher average token frequencies than regular verbs
\citep{Marcus1992}. These two imbalances conspire to make this a
particularly challenging matching problem, though word frequency is an
extremely important covariate for many behavioral tasks.

The inventory of English verbs used here was created by combining of a
list of verbs from the English Lexicon Project \citep{Balota2007}, a
list of irregular verbs extracted from the CELEX-2 lexical database
\citep{Baayen1996}, and word frequencies norms from SUBTLEX-US database
\citep{BrysbaertNew2009}. This data set contains 3,727 verbs, of which
134 are irregular. 

\subsubsection{Matching criteria}
\label{item-matching-criteria}

\autoref{matching.verbs.sbtlx.freq.t_test.0.2} shows
experiments matching regular and irregular verbs on word frequency using
the criterion \(p(t) \ge .2\). A second set of experiments, shown in
\autoref{matching.verbs.sbtlx.freq.sbtlx.basefreq.sbtlx.pformbase.t_test.0.2},
matches these groups on both word frequency as well as the conditional
probability of the verb given its \say{base} (i.e., citation form or
lemma), a measure proposed by \citet{LignosGorman2012}. 

\subsubsection{Results}
\label{item-results}

Each algorithm is applied ten times; the tables display the number of times the
algorithms produce outcomes, and the range of the resulting values. For
both sets of experiments, the only feasible approaches are
\texttt{heuristic3} and \texttt{heuristic4} with limited lookahead. We
also experiment with another method to speed up computation. Let
\(\rho\) denote the frequency with which \(r\), the
\(p\)-value/threshold ratio, is updated. Whereas we use \(\rho = 1\)
above---that is, \(r\) is recomputed every iteration---we also perform
matching with \(\rho = 10\) and \(\rho = 100\). When \(\rho = 10\), for
instance, this produces a considerable speed-up and the resulting
matches are only slightly degraded from the solutions obtained with
\(\rho = 1\). However, high values of $\rho$ do not result in usable matches.

\begin{table}
\centering
\begin{tabular}[c]{@{}lrrrr@{}}
\toprule
algorithm & \# outcomes & \# EI & \# ER & time \\
\midrule
\(h_3\) (\(\lambda = 1\)) & 10 & 49 & 1,060 & 18 hours \\
\(h_4\) (\(\lambda = 1\)) & 10 & 49 & 1,060 & 18 hours \\
\(h_3\) (\(\lambda = 1\), \(\rho = 10\)) & 10 & 49 & 1,060 & 7 hours \\
\(h_4\) (\(\lambda = 1\), \(\rho = 10\)) & 10 & 49 & 1,060 & 7 hours \\
\(h_3\) (\(\lambda = 1\), \(\rho = 100\)) & 10 & 50 & 1,082 & 17 minutes \\
\(h_4\) (\(\lambda = 1\), \(\rho = 100\)) & 10 & 51 & 1,104 & 19 minutes \\
\bottomrule
\end{tabular}
\caption{Results from ten replications of matching regular and irregular
verbs on word frequency at \(\alpha = .2\). \# outcomes: number of
successful runs; \# EI: number of excluded irregular verbs (best of
successful runs); \# ER: number of excluded regular verbs (best of
successful runs); time: minimum wall clock execution time across
successful replications. \(h_3\): \texttt{heuristic3}; \(h_4\):
\texttt{heuristic4}; \(\lambda\): lookahead; \(\rho\): number of items
removed before recomputing item scores.}
\label{matching.verbs.sbtlx.freq.t_test.0.2}
\end{table}

\begin{table}
\centering
\begin{tabular}[c]{@{}rrrrr@{}}
\toprule
algorithm & \# outcomes & \# EI & \# ER & time \\
\midrule
\(h_3\) (\(\lambda = 1\)) & 9 & 63 & 2,811 & 81 hours \\
\(h_4\) (\(\lambda = 1\)) & 10 & 63 & 2,970 & 82 hours \\
\(h_3\) (\(\lambda = 1\), \(\rho = 10\)) & 6 & 63 & 2,598 & 9 hours \\
\(h_4\) (\(\lambda = 1\), \(\rho = 10\)) & 3 & 63 & 3,195 & 9 hours \\
\(h_3\) (\(\lambda = 1\), \(\rho = 100\)) & 8 & 63 & 2,953 & 1 hour \\
\(h_4\) (\(\lambda = 1\), \(\rho = 100\)) & 7 & 63 & 3,005 & 1 hour \\
\bottomrule
\end{tabular}
\caption{Results from ten replications of matching regular and irregular
verbs on word frequency and the conditional probability of word given
morphological base (\texttt{pformbase}) at \(\alpha = .2\). In all
successful experiments, a large number of entries must be removed to
satisfy the criterion; failed experiments are omitted. \# outcomes:
number of successful runs; \# EI: number of excluded irregular verbs
(best of successful runs); \# ER: number of excluded regular verbs (best
of successful runs); time: minimum wall clock execution time across
successful replications. \(h_3\): \texttt{heuristic3}; \(h_4\):
\texttt{heuristic4}; \(\lambda\): lookahead; \(\rho\): maximum number of
items removed before recomputing item scores.}
\label{matching.verbs.sbtlx.freq.sbtlx.basefreq.sbtlx.pformbase.t_test.0.2}
\end{table}

\section{Discussion}
\label{discussion}

The most important difference between the algorithms is running time.
The relatively-simple \texttt{heuristic2} returns a solution in seconds
or minutes for the problems considered here, whereas algorithms with
relatively large look-aheads may require hours or even days of parallel
computation. Nevertheless, in certain cases it may be worth running such
methods for many days. For instance, one may wish to commit a large
amount of compute time simply to preserve one or two more subjects when
there is a substantial marginal cost associated with each subject's
data, and the resulting solution may be used in multiple projects and/or
publications. For instance, both \citet{Kiss2017} and
\citet{MacFarlane2017} use the CSLU ERPA Corpus look-ahead
subject-matches, which required several days of computation.

The \texttt{heuristic3} and \texttt{heuristic4} methods often identify
multiple solutions which are roughly equivalent with respect to the
provided criteria. This can be useful to confirm that findings
generalize across various subsamples matched to the same criteria.
Alternatively, since there is a stochastic element to the proposed
heuristics, replications may result in slightly different outcomes, and
so one may wish to run these heuristics repeatedly times and analyze the
best solution.

While the heuristics we pose here have proved effective for our
purposes, future work could frame the matching problem using integer
linear programs or their the linear relaxations. Alternatively, one
could use \concept{local-improvement algorithms}
\citep{AartsLenstra2003} to refine solutions found via heuristic.

The above experiments used $p$-values from null hypothesis tests as
criteria for matching. However, since our stated goal is to create
groups which were equivalent---not to accept or reject the null
hypothesis---future work could use Bayesian statistical methods which
allow one directly estimate the probability that experimental groups are
\say{practically equivalent} with respect to their covariates
\citep[e.g.,][]{Benavoli2017}.

\section{Conclusions}
\label{conclusions}

We have proposed an approach for matching groups of subjects or items so
as to minimize covariate differences between the groups. Evaluation of
the proposed methods for matching found that even the simplest
algorithms often find acceptable solutions, and these solutions are
improved using methods which are computationally more intensive. The
implementations above are available to the research community in the
form of the R package \texttt{ldamatch}, which can be obtained from the
CRAN repository.

\section*{Acknowledgments}
\label{acknowledgments}

We thank Steve Bedrick for technical assistance with cluster computing.

This material is based on work supported by the National Institute on Deafness and Other Communication Disorders of the National Institutes of Health under awards R01DC007129 and R01DC012033, and by Autism Speaks under Innovative Technology for Autism Grant 2407. The content is solely the responsibility of the authors and does not necessarily represent the official views of the granting agencies or any other individual.

\bibliographystyle{apalike}
\bibliography{matching}

\end{document}